# Exploring Navigation Styles in a FutureLearn MOOC

Lei Shi[1], Alexandra I. Cristea[1], Armando M. Toda[2], Wilk Oliveira[2]

[1] Durham University, Durham, UK
{lei.shi, alexandra.i.cristea}@durham.ac.uk
[2] University of São Paulo, São Paulo, Brazil
{armando.toda, wilk. oliveira}@usp.br

**Abstract.** This paper presents for the first time a detailed analysis of fine-grained navigation style identification in MOOCs backed by a large number of active learners. The result shows 1) whilst the sequential style is clearly in evidence, the global style is less prominent; 2) the majority of the learners do not belong to either category; 3) navigation styles are not as stable as believed in the literature; and 4) learners can, and do, swap between navigation styles with detrimental effects. The approach is promising, as it provides insight into online learners' temporal engagement, as well as a tool to identify vulnerable learners, which potentially benefit personalised interventions (from teachers or automatic help) in Intelligent Tutoring Systems (ITS).

## 1   Introduction

Having emerged in recent years, especially with the rise of big data, Learning Analytics (LA) [1] hugely impacts on the area of Intelligent Tutoring Systems (ITS). Common goals include predicting performance and retention, as well as improving assessment and engagement [2, 3]. Effective LA practice often involves statistical modelling for meaningful insights. The ever-increasing amount of learner data from MOOCs (Massive Open Online Courses) brings unprecedented opportunities to enhance LA. In turn, LA provides methods to determine factors influencing learner behavior, allowing improvements of learning context, design and pedagogies [4].

   Patterns in general have fascinated humankind [5]. Learning patterns have been studied for a long time, both offline and, since the advent of the Internet, online [6]. Navigation Styles is more recently of interest, especially related to self-directed learning in MOOCs, placing the control with the learners. This recent interest brings insights into data-driven re-examination of traditional theories of Learning Styles. MOOCs are notoriously known for low completion [7], so we aim to re-examine in depth actual learner behavior and understand how to better help, by answering 4 cumulative **research questions**: *RQ1 what are the real navigation styles of MOOC learners? RQ2 how do these navigation styles relate to traditional theories of Learning Styles? RQ3 how do different navigation styles affect the course* *completion?* ***RQ4*** *which are the learners that are particularly in need of help?*

## 2   Related Work

Many studies have been conducted using Learning Analytics (LA) to understand learner behavior thus improve online learning engagement. For example, Pardo *et al.* [8] addressed the challenges impeding the capacity of instructors to provide personalized feedback at scale. Zhang *et al.* [9] explored the role of Slack in collaborative learning engagement. Shoufan [10] investigated how YouTube videos could support online learning. Cristea et al. [11] predicted dropout for earlier interventions. Shi *et al*. [12] analyzed behavioral and demographic patterns. Most of these studies grouped learners and compared patterns amongst the groups, aiming at a deeper understanding of how online learners engage and perform. Our study, presented in this paper, also takes the grouping approach, but, differently, we group learners based on how they follow the *pre-defined directed linear learning path*.

Time is a fundamental dimension of learning. When breaking down a course into temporal phases, existing relationships amongst various parameters or variables may not continue along the learning process [13]. Some studies have taken into consideration this temporal dimension. For example, Yang *et al.* [14] clustered learners based on their characteristics and their interactions with learning materials to understand how their cluster membership changed along the course. Laer and Elen [15] examined changes in learner behavior and outcomes to test if providing cures for calibration affects self-regulated learning. In the current study, differently, we analyze how learners navigate in the *pre-defined directed linear learning path*, group them using this information, and compare their engagement and performance.

The effects of learning style models including the well-known Felder-Silverman [16] and Kolb [17] models have been well studied. Despite criticism on the concept of learning styles [18], many studies e.g. [19] incorporated the concept to support personalized learning, and claimed their findings strongly approve the concept. In our study, we group learners by navigation styles rather than learning styles and investigate how the identified navigation styles relate to theories of learning styles. We map navigation styles to the *sequential-global* dimension of Felder-Silverman Learning Styles [16], which determines how learners prefer to progress toward understanding information. *Sequential learners* prefer linear thinking and learn in small incremental steps, whereas *global learners* prefer holistic thinking and learn in large leaps. Note that sequential and global styles exist on a continuum – with some learners heavily favoring one or the other, and others using a little bit of both.

## 3   Experimental Settings

The dataset was taken from "The Mind is Flat", a 6-week course on the FutureLearn MOOCs platform. Note that, only one course was chosen, because learning styles rely on subject of learning thus combining courses may cancel out the effect of the navigation style identification. Each week contained several steps – basic learning units which could be an article, a discussion forum, or an assessment. There were {14,12,14,12,12,18} steps in {Week 1, …, Week 6}. Each week had an Introduction step, 8~12 "Main" steps (articles or videos, the main learning content), a Discussion step, an Assessment step, and a Review step. Week 6 contained 2 additional steps: a "Further Reading" step, recommending academic papers to read after the course, and a "Certificates" step, promoting buying a printed statement of participation. The *pre-defined linear learning path* for learners to follow each week was to start with the Introduction step, then to visit the "Main" step(s), participate in the Discussion step, visit the Experiment and Assessment steps, and finish with the Review step. Initially, 14,240 learners enrolled in the course, with 2,030 unenrolled, thus 14,240-2,030=12,210 remaining. FutureLearn defines active learner as "a learner who goes on to mark at least one step as complete in a course" [20]. Using this definition, our dataset had 5,204 active learners. As the rest 12,210-5,204=7,006 learners barely did anything, our analysis focused on only the activity logs of 5,204 active learners.

We first inspected Week 1, identifying navigation styles and exploring whether and how these styles might correlate to engagement and performance. *Engagement* was measured by how learners accessed learning content, e.g., reading articles, and how they participated in discussion forums and took assessments. *Performance* was measured via the assessment results (correct answer rate). Note that, as all the weeks in the course were structured the same, navigation styles can be identified at the week-level and explored for stability and persistence in further weeks. Moreover, identifying navigation styles early in Week 1 potentially allows early intervention and help for learners. Hence, we started by analyzing **Week 1** (section 4.1). We then moved onto the analysis of whether and how navigation styles might change **across weeks** (section 4.2). The temporal analysis aimed to thus understand temporal stability of the learning process and analyze consequences of the style choice.

## 4    Results and Discussions

### 4.1    Week 1

In Week 1, there were 14 steps including 1 Introduction step, 9 Main steps, 1 Discussion step, 1 Experiment step, 1 Assessment step, and 1 Review step.

   *How did learners start?* The majority (90.3%; 4,699/5,204) active learners started with the Introduction step. Among the rest 506 learners, 438 (86.6%) started with a Main step; 19 (3.8%) the Experiment step, 2 (0.4%) the Discussion step, 29 (5.7%) the Assessment step, and 18 (3.6%) the Review step. Despite 506 learners not starting with the Introduction step, 96.0% (4,998/5,204) of them did visit it (after visiting one or a few other steps). This implies the perceived importance of the Introduction step, despite learners' navigation styles. Indeed, it introduced both what and how to learn – helpful in providing learners with the 'big picture' of the course, including what to expect now and later, and how to proceed.

   *How did learners follow the pre-defined directed linear learning path?* 5,135 (98.7% of 5,204) active learners visited at least one Main step. Fig. 1 shows the number of Main steps visited by the learners in Week 1, with only 72 (1.4%) learners not visiting any Main step, and 2,286 (43.9%) visiting all (9). 2,172 (41.7% of 5,204) active learners visited the Discussion step; and surprisingly, 99.7% (2,165/2,172) participated in the discussion. This indicates that Discussion, in itself, is very important to learners in the course, and it might generally be important to learners online, as they might come to online courses partially for social reasons.

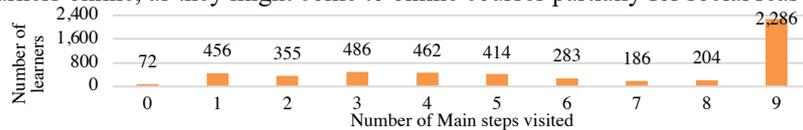

**Fig. 1.** Distribution of Frequency of accessing Main *steps*

Whilst the Discussion step was clearly designed to be visited after having visited all the Main steps (the *pre-defined directed linear learning path*, as explained in section 3), 1.6% (83/5,204) active learners visited one or a few Main steps *after* the Discussion step. These learners might be in the online course for discussion only. 2.9% (153/5,204) visited some Main step(s) after the Experiment step. 4.4% (229/5,204) visited some Main step(s) after the Assessment step. These learners might have actively been trying to use their current knowledge to test themselves, before starting learning, or might have wanted to check how challenging the tests were, before committing to learning. The fact that they returned to learning (the Main steps) afterwards seems to suggest their current knowledge was not (yet) sufficient to pass. 1.9% (98/5,204) visited some Main step(s) after the Review step. They might have wanted to know the main goals of the course before committing to learning, for a holistic/global view. Nonetheless, it is interesting that the majority of the learners conformed with the linear expectations on visiting the Main step first.

Fig. 2 shows the average number of Main steps visited before (blue columns) and after (red columns) visiting the Discussion, Experiment and Assessment steps, respectively. Overall, learners visited more Main steps before one of these special steps, which they should have visited after visiting *all* the Main steps. Yet, there was still a large percent of Main steps visited afterwards. This shows more learners than the above extreme ones felt the need to have an overview, to know the learning goals, or to see if the Experiment and/or the Assessment was something for them. it also shows, in average at least, learners tended to be non-linear, as defined below.

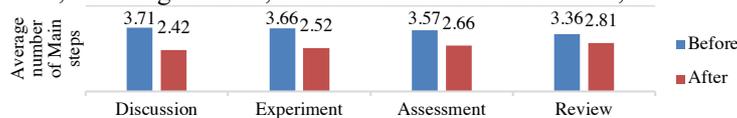

**Fig. 2.** The average number of Main steps visited BEFORE and AFTER visiting other *steps*

We define **Sequential** learners as those active learners who strictly followed the pre-defined directed linear learning path, rendered closest to Felder-Silverman's "sequential" – one extreme end of the sequential-global continuum [16]. Among 5,204 active learners, 1,552 (29.8%) were Sequential. We define **Global** learners as those active learners who visited at least one Main step for the first time after either the Discussion, the Experiment, the Assessment, or the Review steps, which were successors of *all* the Main steps in the learning path, and we map this learner group over Felder-Silverman's "global" – the other extreme end of the sequential-global continuum [16]. There were 350 (6.7% of 5,204) Global learners. The rest of the active learners' navigation styles didn't comply with either extreme end but fell in between. We thus define them as **Middle** learners. Next, we compare engagement and performance (as introduced in section 3) among these three learner groups.

*How did learners participate in Discussion and Assessment steps?* Sequential learners had the highest mean value for both comments ($\mu=0.3$) and attempts ($\mu=13.6$), followed by Global learners ($\mu_{comments}=0.22$; $\mu_{attempts}=9.1$); while Middle learners had the lowest mean value for both comments ($\mu=0.02$) and attempts ($\mu=0.8$). It means Sequential learners, who followed the course in the prescribed manner, might also be the most dedicated ones, whereas Global learners might be wasting time going back and forth thus missing activities. Middle learners seemed to be the least engaged. This, together with the fact that these learners represent the great majority, may not be surprising in the MOOCs context – it is a known fact in such e-learning systems, the majority of learners do drop out from the course [21].

Fig. 3 (left) compares the distribution of comments amongst these three learner groups on a logarithmic scale (base=10). They share a similar distribution pattern: the majority did not leave any comment, and most of them left very few. However, it is obvious from the image that Sequential learners commented the most, followed by Global learners; whilst no Middle learner commented more than 2 comments. It is interesting that, whilst Sequential learners were indeed, again, the most active, Global learners were only marginally more active than Middle learners in terms of commenting (social). This may show a different focus of Global learners, less on social aspects, prioritizing the learning. Fig. 3 (right) compares the distribution of attempts made in the Assessment step on a logarithmic scale (base=10), showing a similar pattern, yet many more Sequential learners made much more attempts. A high number of learners in each group did not do any attempt. For the rest, the distribution is almost Gaussian, with peaks between 12-14 attempts for all groups.

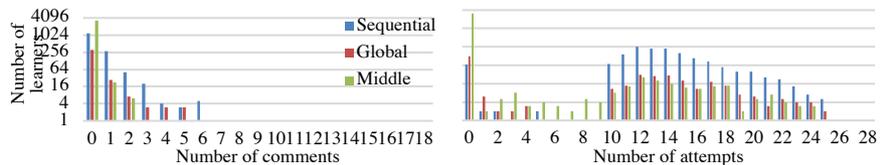

**Fig. 3.** Distribution of comments (left) and attempts (right)

The result from a Kolmogorov-Smirnov test indicated a non-normal distribution for comments in Week 1 (Discussion step; $D(5,204)=0.502$, *p<0.001*); this is also the case for the attempts to answer the questions in the Assessment step ($D(5,204) = 0.399$, *p<0.001*). Thus, the nonparametric Kruskal-Wallis H and Mann-Whitney U tests were used for the comparisons amongst these three learner groups. The result from a Kruskal-Wallis H test showed statistically significant differences in both comments ($\chi^2(2)=681.33$, *p<0.001*) and attempts ($\chi^2(2)=3,579.24$, *p<0.001*). Mann-Whitney U tests for pairwise comparisons were conducted, showing both comments ($Z=-4.13$, $U=144,385.0$, *p<0.001*) and attempts ($Z=-8.71$, $U=191,226.0$, *p<0.001*) were significantly higher for Sequential than for Global learners; both comments ($Z=-14.602$, $U=510,566.00$, *p<0.001*) and attempts ($Z=-32.498$, $U=241,255.00$,

*p<0.001*) being significantly higher for Global than for Middle learners.

***How did learners perform in the Assessment?*** Overall, Sequential learners had at least one attempt to answer the questions in the Assessment step; whereas that number for Global and Middle learners was of 223 (63.7% of 350) and 194 (5.9% of 3,302), respectively. Here, only learners who had at least one attempt were considered: 1,485+223+194=1,885 learners. The *correct answer rate* was defined as the number of questions correctly answered, divided by the number of questions attempted. Sequential learners had the highest mean and median for *correct answer rate* ($\mu$=70.0%, $\sigma$=16.8%,, $\tilde{x}$=71.4%), followed by Global learners ($\mu$=64.3%, $\sigma$=21.8%,, $\tilde{x}$=66.7%) and then Middle learners ($\mu$=59.3%, $\sigma$=23.1%,, $\tilde{x}$=61.3%).

Fig. 4 shows the distribution of the *correct answer rates* amongst the three learner groups. In general, at a higher *correct answer rate* category, i.e. [60%,70%), [70%,80%), [80%,90%), and [90%,100%], the proportion of Sequential learners is the highest, followed by Global learners. At the low end, only 0.2% (3/1,458) of Sequential learners didn't answer any question correctly, with a much larger 3.1% (7/223) proportion of Global and 3.6% (7/194) Middle learners. Interestingly, the proportion of Global learners who correctly answered all (100%) questions attempted was the highest, 5.4% (12/223), in comparison with Sequential learners (4.5%; 65/1,458) and Middle learners (4.1%, 8/194). This may be because some percentage (although small) of Global learners may have had prior knowledge, although chance may play a role (as the result is not statistically significant).

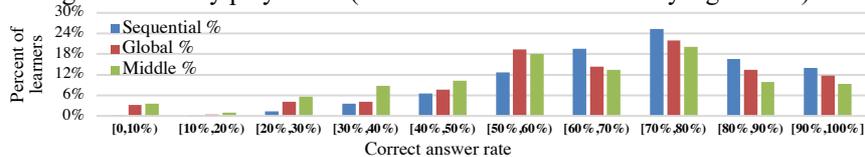

**Fig. 4.** Distribution of correct answer rate

The Kolmogorov-Smirnov test result indicated a non-normal distribution of the *correct answer rates*, D(1,902)=0.10, *p<0.001*, so the Kruskal-Wallis H and Mann-Whitney U tests were used to compare the *correct answer rate* for learner groups. The Kruskal-Wallis H test result suggested a statistically significant difference amongst these three learner groups in *correct answer rate*, $\chi 2(2)$=44.7, *p<0.001*. The Mann-Whitney U test result suggested *correct answer rates* to be significantly higher for Sequential learners than for both Global and Middle learners.

***How did learners finish in Week 1?*** As stated in Section 3, the Review *step* was the end of the *pre-defined directed linear learning path*. Among the 5,204 active learners, 4,331 (83.22%) visited it, but only 1,847 (35.49% of 5,204) finished the current week with it. For the rest, 84 (1.61% of 5,204) learners finished with the Introduction, 2,864 (55.03%) with a Main, 97 (1.86%) with the Discussion, 132 (2.54%) with the Experiment, and 180 (3.46%) with the Assessment. Comparing learning groups, in terms of if the last step the learners visited in Week 1 was the Review, whilst, by definition, 100% Sequential learners did visit the Review step as the last step, there were only 39.43% (138/350) for Global and 4.8% (157/3302) for Middle learners. This, again, suggests Middle learners were the least active.

### 4.2 Across-Weeks

***Changing navigation style.*** As presented in section 4.1, in Week 1, there were 1,552 (29.82% of 5,204) Sequential, 350 (6.73%) Global and 3,302 (63.45%) Middle learners. These numbers and proportions changed however in Weeks 2 to 6 (Fig. 5). Fewer and fewer learners behaved as Sequential; more and more as Middle; whereas Global fluctuated – the peek appeared in Week 2 and the smallest number appeared in Week 5 (neither the first nor last week). This could be due to the fact that learners who enrolled had less time and were trying to see beforehand if they could perform the final activities in weeks, to finish potentially earlier. The fact that more learners acted as Middle learners is, however, a more worrying trend, and is potentially an early indicator for learners hovering on the brink of leaving the course altogether.

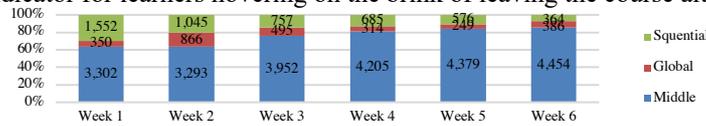

**Fig. 5.** Learner groups changes over weeks

Fig. 6 shows how learners switched from one group to another, i.e. the transition probabilities between ***navigation styles***. The strength of the arrows represents the proportion of learners switching; e.g., from Weeks 1 to 2, 57.15% (887) Sequential learners remained Sequential, whilst 29.45% (457) became Global and 13.40% (208) became Middle. **Across weeks**, the majority (57.15%~77.15%) of Sequential learners remained i.e. kept their navigation style; yet, for Global learners, the largest percentage (41.42%~67.55%) became Middle, 21.02%~39.76% remained Global, and only 8.03% ~ 20.29% became Sequential. The Middle group is the most stable, with 89.04%~98.86% remaining, and the smallest percent (0.16%~2.63%) becoming Sequential. Interestingly there is less of a "loss" between Sequential and Global learner groups, when compared to the "loss" between Global and Middle.

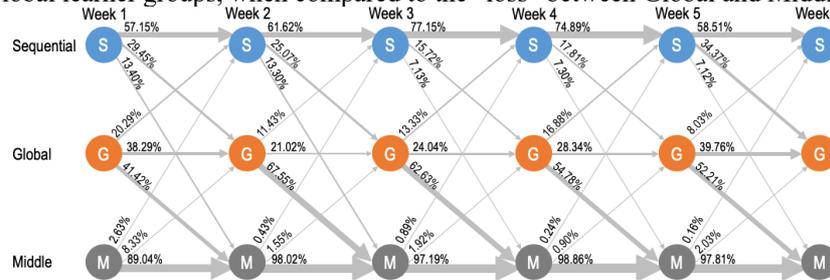

**Fig. 6.** Learners switching between groups

Each learner's *Navigation style* in each week was labelled: S for Sequential, G for Global, and M for Middle, resulting in 236 distinct *navigation style changing paths*; e.g. "SGSSGM" refers to learners who were Sequential in Week 1 and became Global in Week 2, etc. In total, 3,076 (59.11%) learners kept navigation style, i.e. remaining "SSSSSS", "GGGGGG" or "MMMMMM"; whereas 2,128 (40.89%) learners changed navigation style at least once. The majority (2,824;

54.27%) kept being Middle i.e. "MMMMMM". They were the least active: on average only visiting 5.50 (σ=2.90) and completing 4.20 (σ=2.74) steps. Another 252 learners also kept their navigation style across weeks: 234 "SSSSSS" and 18 "GGGGGG" learners; the former visited 82 (σ=0) and completed 81.17 (σ=5.04) steps, and the latter visited 75.39 (σ=12.40) and completed 70.22 (σ=22.26) steps. Other popular navigation style changing paths include "SGMMMM" for 286 learners, "SMMMMM" for 192 learners, and "MGMMMM" for 177 learners.

*__Completion.__* Fig. 7 shows the distribution of the steps that learners stopped the course. For Sequential and Global groups, peaks appear towards the end of the course, meaning the largest proportion of learners in these two groups stopped after having visited some part of the course. In particular, 23.07% (358/1,552) Sequential learners stopped after having visited the last step - they completed the whole course; whilst 8.29% (29/350) Global learners completed the whole course. The majority of the Sequential learners (87.82%; 1,363/1,552), and more than half of the Global learners (61.71%; 216/350) went on visiting step(s) in Week 2 and later weeks. On the contrary, the largest proportions of Middle learners stopped in Week 1: 12.78% (422/3302) stopped at Step 1.2, and {9.06%,12.75%,12.11%,10.84%} stopped at Step {1.3,1.4,1.5,1.6}; and only 14.51% of them went on visiting other step(s).

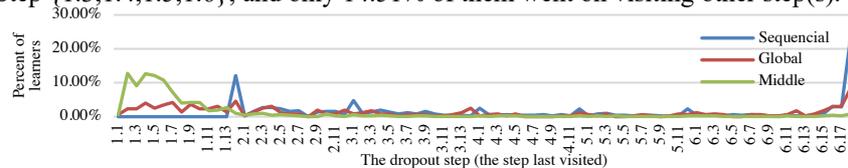

**Fig. 7.** Distribution of dropout steps

Interestingly "peaks" for Sequential learners appear periodically at the end of each week. This means if a Sequential learner were to stop, they likely stopped after visiting all steps in the current week; but, for Global and Middle learners, there isn't a clear periodical pattern of such – in line with the sequential learning style (compared to the global learning style [16]), where learners who have this preference would complete one subject at the time. Presumably, they don't like to stop in the middle of a subject but have the need (or compulsion) to finish it first, before they either move on to a new subject, or to a different course/endeavor.

## 5    Conclusions

We've analyzed in-depth learners' navigation styles on a FutureLearn MOOC with massive learner presence. We've also investigated the changes of navigation styles along the course. The results and discussions have answered four research questions raised in Section 1. For **RQ 1** what are the real navigation styles of MOOC learners? we've identified 3 navigation styles including Sequential, Global and Middle, based on whether and how the learners followed the pre-defined directed linear learning path. For **RQ 2** how do these navigation styles relate to traditional theories of Learning Styles? we've mapped the 3 identified navigation styles over the well-known Felder-Silverman Learning Styles [16]. In particular, Sequential and Global are mapped over the 2 extreme ends of the sequential-global continuum, and the third style, i.e. Middle, represents the learners whose navigation style did not comply with either extreme end but fell in between. For **RQ 3** how do different navigation styles affect the course completion? we've shown the navigation style chosen is strongly related to the chance of completion, and the variability in styles is related to this outcome, too. For **RQ 4** which are the learners that are particularly in need of help? we've identified learners in need for help, especially amongst those that have potentially better chances of completion e.g. Global learners.

Overall, this study shows clearly navigation styles are important for the success in MOOCs, and they can be related with dropout/completion. However, this study also shows styles aren't constant for learners, and, whilst they might prefer studying in a specific style in general, swaps between styles still can happen relatively often. We show which styles are more "stable" in this sense than others. Unfortunately, the most stable style, Middle, is also the one most likely to lead to disengagement and ultimately leaving the course. The Sequential style is the most successful, but still has a great loss towards other styles. Global is the most unstable style – such learners still have a chance to succeed, in terms of completing either the whole or a good proportion of the course, but they clearly need help in remaining on track and being focused. This brings us to the other main finding, that of finding potentially the most vulnerable category of learners, identifiable by their navigation style early on starting from week 1. Therefore, online teachers or personalized systems can focus on helping them. On the other hand, the Middle style is broad and mostly unsuccessful - in need of further analysis in terms of refining the type of behavior encountered within it, to provide specialized and personalized help and direction.

## References

1. 1st International Conference on Learning Analytics and Knowledge 2011 | Connecting the Technical, Pedagogical, and Social Dimensions of Learning Analytics, https://tekri.athabascau.ca/analytics/, last accessed 2020/03/01.
2. Shi, L., Cristea, A.I.: In-depth Exploration of Engagement Patterns in MOOCs. In: Hacid, H., Cellary, W., Wang, H., Paik, H.-Y., and Zhou, R. (eds.) Web Information Systems Engineering – WISE 2018. pp. 395–409. Springer International Publishing, Cham (2018). https://doi.org/10.1007/978-3-030-02925-8_28.
3. Zacharoula Papamitsiou, Anastasios A. Economides: Learning Analytics and Educational Data Mining in Practice: A Systematic Literature Review of Empirical Evidence. Journal of Educational Technology & Society. 17, 49–64 (2014).
4. Ferguson, R., Clow, D.: Examining Engagement: Analysing Learner Subpopulations in Massive Open Online Courses (MOOCs). In: Proceedings of the Fifth International Conference on Learning Analytics And Knowledge - LAK '15. pp. 51–58. ACM Press, Poughkeepsie, New York (2015). https://doi.org/10.1145/2723576.2723606.
5. Alexander, C.: A Pattern Language: Towns, Buildings, Construction. OUP USA, New York (1978).
6. Romero, C., Ventura, S.: Educational Data Mining: A Survey from 1995 to 2005. Expert

Systems with Applications. 33, 135–146 (2007). https://doi.org/10.1016/j.eswa.2006.04.005.
7. Alamri, A., Alshehri, M., Cristea, A., Pereira, F.D., Oliveira, E., Shi, L., Stewart, C.: Predicting MOOCs Dropout Using Only Two Easily Obtainable Features from the First Week's Activities. In: Coy, A., Hayashi, Y., and Chang, M. (eds.) Intelligent Tutoring Systems. pp. 163–173. Springer International Publishing, Cham (2019). https://doi.org/10.1007/978-3-030-22244-4_20.
8. Pardo, A., Jovanovic, J., Dawson, S., Gašević, D., Mirriahi, N.: Using Learning Analytics to Scale the Provision of Personalised Feedback. Br J Educ Technol. 50, 128–138 (2019). https://doi.org/10.1111/bjet.12592.
9. Zhang, X., Meng, Y., Ordóñez de Pablos, P., Sun, Y.: Learning analytics in collaborative learning supported by Slack: From the perspective of engagement. Computers in Human Behavior. 92, 625–633 (2019). https://doi.org/10.1016/j.chb.2017.08.012.
10. Shoufan, A.: Estimating the cognitive value of YouTube's educational videos: A learning analytics approach. Computers in Human Behavior. 92, 450–458 (2019). https://doi.org/10.1016/j.chb.2018.03.036.
11. Cristea, A.I., Alamri, A., Kayama, M., Stewart, C., Alshehri, M., Shi, L.: Earliest Predictor of Dropout in MOOCs: A Longitudinal Study of FutureLearn Courses. Presented at the 27th International Conference on Information Systems Development (ISD2018), Lund, Sweden August 22 (2018).
12. Shi, L., Cristea, A., Toda, A., Oliveira, W.: Revealing the Hidden Patterns: A Comparative Study on Profiling Subpopulations of MOOC Students. In: The 28th International Conference on Information Systems Development (ISD2019). Association for Information Systems, Toulon, France (2019).
13. Zhu, M., Bergner, Y., Zhang, Y., Baker, R., Wang, Y., Paquette, L.: Longitudinal Engagement, Performance, and Social connectivity: A MOOC Case Study using Exponential Random Graph Models. In: Proceedings of the Sixth International Conference on Learning Analytics & Knowledge - LAK '16. pp. 223–230. ACM Press, Edinburgh, United Kingdom (2016). https://doi.org/10.1145/2883851.2883934.
14. Yang, B., Shi, L., Toda, A.: Demographical Changes of Student Subgroups in MOOCs: Towards Predicting At-Risk Students. Presented at the The 28th International Conference on Information Systems Development (ISD2019), Toulon, France August (2019).
15. Van Laer, S., Elen, J.: The effect of cues for calibration on learners' self-regulated learning through changes in learners' learning behaviour and outcomes. Computers & Education. 135, 30–48 (2019). https://doi.org/10.1016/j.compedu.2019.02.016.
16. Felder, R.M., Silverman, L.K.: Learning and Teaching Styles in Engineering Education. 78, 674–681 (1988).
17. Kolb, A.Y., Kolb, D.A.: Learning Styles and Learning Spaces: Enhancing Experiential Learning in Higher Education. Academy of Management Learning & Education. 4, 193–212 (2005).
18. Kirschner, P.A.: Stop propagating the learning styles myth. Computers & Education. 106, 166–171 (2017). https://doi.org/10.1016/j.compedu.2016.12.006.
19. Hassan, M.A., Habiba, U., Majeed, F., Shoaib, M.: Adaptive gamification in e-learning based on students' learning styles. Interactive Learning Environments. 0, 1–21 (2019). https://doi.org/10.1080/10494820.2019.1588745.
20. O'Grady, N.: Are Learners Learning? (and How do We Know?), https://about.futurelearn.com/research-insights/learners-learning-know, last accessed 2019/02/23.
21. Clow, D.: MOOCs and the Funnel of Participation. In: Proceedings of the Third International Conference on Learning Analytics and Knowledge - LAK '13. p. 185. ACM Press, Leuven, Belgium (2013). https://doi.org/10.1145/2460296.2460332.